\documentclass[preprint,showpacs,showkeywords,amsmath,amssymb,12]{revtex4}
\usepackage{graphicx,color,float,flafter,epsfig,bm,textcomp,url}

\newcommand{\lcap}{l_{\mathrm{capillary}}}
\newcommand{\ucap}{u_{\mathrm{capillary}}}

\newcommand{\pL}{p_{\mathrm{Laplace}}}
\newcommand{\tcap}{t_{\mathrm{capillary}}}
\newcommand{\that}{\hat{t}}
\newcommand{\psib}{{\psi_{\mathrm{b}}}}
\newcommand{\psis}{\ensuremath{\psi_{\mathrm{s}}}}
\newcommand{\bnabla}{\ensuremath{\vec{\nabla}}}

\newcommand{\fieq}{{f^{\mathrm{eq}}_{i}}}

\newcommand{\bci}{{\vec{c}_i}}
\newcommand{\bfdriv}{{\vec{f}^{\mathrm{driv}}_{\mathrm{Laplace}}}}
\newcommand{\fxdriv}{{f^{\mathrm{driv}}_{x,\mathrm{Laplace}}}}
\newcommand{\bc}{{\vec{c}}}
\newcommand{\bu}{{\vec{u}}}

\newcommand{\bs}{{\vec{s}}}
\newcommand{\br}{{\vec{r}}}
\newcommand{\nuT}{{\nu_{T}}}
\newcommand{\nurho}{{\nu_{\rho}}}
\newcommand{\rhoc}{{\rho_{\mathrm c}}}
\newcommand{\Tc}{{T_{\mathrm c}}}
\newcommand{\eref}[1]{{Eq.\ (\ref{#1})}}
\newcommand{\erefs}[2]{{Eqs.\ (\ref{#1}) and (\ref{#2})}}
\newcommand{\fref}[1]{{Fig.\ \ref{#1}}}
\newcommand{\pc}{{p_{\mathrm c}}}

\newcommand{\rhos}{{\rho_{\mathrm s}}}
\newcommand{\rhoG}{{\rho_{\mathrm Gas}}}
\newcommand{\rhoL}{{\rho_{\mathrm Liquid}}}

\definecolor{red4}{rgb}{0.6,0,0}
\definecolor{green4}{rgb}{0,0.6,0}
\definecolor{green6}{rgb}{0,0.4,0}
\definecolor{blue4}{rgb}{0,0,0.6}
\definecolor{bluegreen4}{rgb}{0,0.6,0.6}

\newcommand{\blue}[1]{{#1}}

\newcommand{\hide}[1]{{#1}}

\begin{document}

\title{Wetting gradient induced separation of emulsions: A combined experimental and lattice Boltzmann computer simulation study}
\author{Fathollah Varnik$^1$, Pagra Truman$^2$, Bin Wu$^1$, Petra Uhlmann$^2$, Dierk  Raabe$^1$ and Manfred Stamm$^2$}
\address{$^1$Max-Planck-Institut f\"ur Eisenforschung, Max-Planck Stra{\ss}e 1, 40237 D\"usseldorf, Germany\\
$^2$Leibniz-Institut f\"ur Polymerforschung, Hohe Stra{\ss}e 6, 01069 Dresden, Germany
}

\date{\today}%

\begin{abstract}
Guided motion of emulsions is studied via combined
experimental and theoretical investigations. The focus of the work
is on basic issues related to driving forces generated via a
step-wise (abrupt) change in wetting properties of the substrate
along a given spatial direction. Experiments on binary emulsions
 unambiguously show that selective wettability of the one of
the fluid components (water in our experiments) with respect to the
two different parts of the substrate is sufficient in order to drive
the separation process. These studies are accompanied by approximate
analytic arguments as well as lattice Boltzmann computer
simulations, focusing on effects of a wetting gradient on internal
droplet dynamics as well as its relative strength compared to
volumetric forces driving the fluid flow. These theoretical
investigations show qualitatively different dependence of wetting
gradient induced forces on contact angle and liquid volume in the
case of an open substrate as opposed to a planar channel. In
particular, for the parameter range of our experiments, slit
geometry is found to give rise to considerably higher separation
forces as compared to open substrate.
\end{abstract}

\pacs{47.11.-j,47.11.Qr,47.85.Np}

\keywords{lattice Boltzmann, microfluidics, separation phenomenon, demixing, capillarity, surface tension}

\maketitle

\section{introduction}
\label{section:introduction}
Guided motion of liquid droplets along the flow in general and
separation of emulsions into their individual components in
particular are important issues in many applications such as
filtration of oil-residuals from water \cite{Holt2006,Faibish2001}
or \blue{separation of protein containing emulsions after initial 
protein separation by liquid-liquid partitioning.}
Individual droplets can also
serve as highly efficient microreactors in order to synthesize e.g.\
nanoparticles and quantum dots \cite{Khan2004} or they can be used in
information processing as bits \cite{Fuerstman2007,Prakash2007}.

Due to the high surface to volume ratio in microfluidic systems,
geometrical confinement to guide liquid flow may be replaced in
parts by chemically patterned surfaces \cite{Zhao2001,Guenther2004}.
Moreover, surface energy gradients can be used to drive liquid
motion.

The surface energy gradients are often generated by external
stimuli, e.g.\ UV-illumination \cite{Ichimura2000}, electrochemical
reactions \cite{Yamada2005} or by a surfactant source
\cite{Lee2005}. These and other experiments demonstrate the strong
influence of surface wettability on liquid dynamics in microfluidic
systems.

In the present work, we use surfaces with 
\blue{an abrupt (step-like) change in wetting properties}
to separate emulsions. The approach based on the use of
a substrate with spatially varying wetting properties (a 'wetting
gradient') is particularly interesting, since it does not require
any additional energy source other than that needed for the
preparation of the chemical gradient.

First experimental studies of wetting gradient induced separation
phenomena were performed by Ionov, Stamm and coworkers on polymer
brush substrates with a continuous variation of the contact angle
\cite{Ionov2006}.

In these studies, the spatial variation of the wetting properties is
controlled by a corresponding variation of the composition of the
individual polymer components. Flow experiments on these substrates
clearly demonstrate the physical significance of the underlying
concept, namely that a spatial variation in wetting properties of
the substrate can lead to the separation of a binary emulsion into its
individual components \cite{Ionov2006}.

However, it is also noted that the system with a continuous wetting
gradient \blue{(as compared to a step-like change in wetting properties 
used in this work)} is too complex to allow a clear identification of
underlying mechanisms leading to a separation of the model emulsion.

\blue{Here, the term 'complex' refers to both the experimental production and the
    separation process itself. (1) Experimental Production: Wettability
    gradients used in previous work have been fabricated using binary polymer
    brushes. This procedure is very sensitive to the experimental conditions
    and the quality of the source material obtained from the
    manufacturer. Furthermore three fabrication steps are required, namely the
    deposition of an adhesive layer, the deposition of the first brush
    component and the deposition of the second brush component. Further
    details can be found in Ref.\ \cite{Ionov2006}. (2) Separation Process:
    Firstly, a step gradient of wettability affects the liquid only locally but
    the separation efficiency increases with the gradient strength which is
    strongest in the case of a step gradient. Secondly a weaker gradient strength
    cannot be compensated by an increasing gradient length due to surface
    energy hysteresis. Thus at the present level of knowledge the step
    gradient of wettability seems to be favorable.}

\blue{In addition to these aspects, the polymer brush gradient
surfaces, which are often used to produce continuous wetting gradients,
are very sensitive to pollution and their fabrication is a
complex time consuming procedure.}

Therefore, there is a strong need to conduct separation experiments
on easy to fabricate, robust, reproducible and well describable
model gradient surfaces. A further, very important aspect is the
need for a theoretical approach allowing not only to rationalize the
experimental findings, but also to study the basic underlying
mechanisms independently. The present work provides such a combined
experimental/theoretical study.

\section{Methods}
\label{section:methods}
\subsection{Experimental Methods}
\subsubsection{Generation of quasi-monodisperse emulsions.}
An important prerequisite for the realization of reproducible
separation experiments is the generation of emulsions with well
defined droplet size. For this purpose, a channel with a diameter of
1.5mm with two in- and one outlet is used to generate a binary emulsion 
(see \fref{fig:emulsion-generation}b  for a
schematic view). Into one inlet the solvent (here toluene) and into
the other inlet water is injected.

As shown in \fref{fig:emulsion-generation}a, the size of the
water and the solvent droplets can be controlled via a variation of
the ratio between the injection rates of the solvent and the water
components. There is, however, a lower limit to the size of water
droplets. The smallest droplet size is determined by the
cross-sectional area of the channel and is approximately 2.9$\mu$l
(micro liter) in this setup. Smaller droplets are 'unstable' in the
sense that they float in the surrounding liquid medium (toluene)
until they come in contact and coalesce, whereby forming larger
droplets. This process goes on until the resulting droplet fills the
channel's cross section. Two such (large) droplets
are then separated along the channel by a drop of toluene thus
preventing their coalescence.

\hide{
\begin{figure}
\epsfig{file=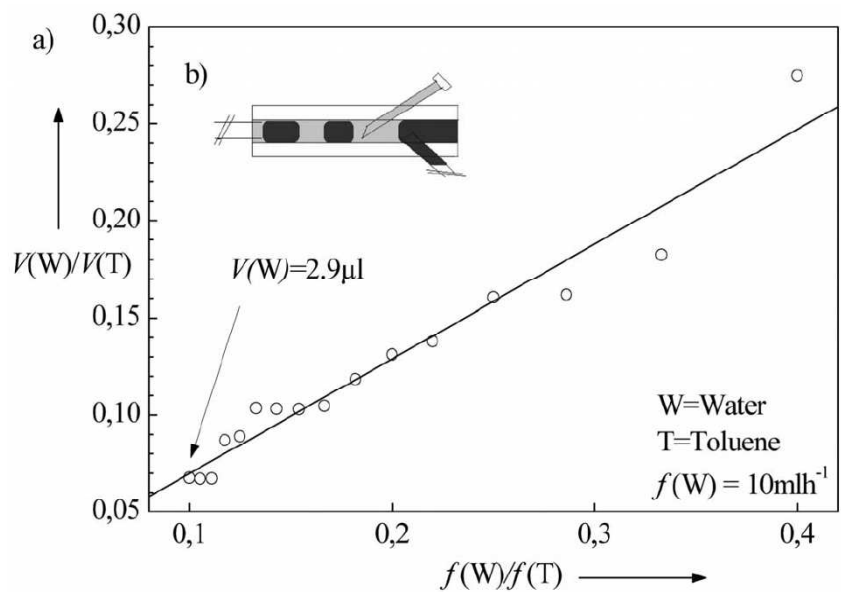,height=80mm,clip=}
\caption[]{a)
Generation of a monodisperse water-toluene mixture using a narrow channel
(b). The horizontal axis is the ratio of water/toluene pump rates. The
vertical axis is the resulting water droplet/toluene volume ratio. An arrow shows the
smallest droplet volume which has been generated via this experimental
setup. This minimum droplet size scales with the wall-to-wall
separation (channel height).
}
\label{fig:emulsion-generation}
\end{figure}}

\subsubsection{Preparation of step-gradient surfaces}
We use, in this series of experiments, the simplest kind of a
wetting gradient, namely a step-wise change in the contact angle. As
shown in \fref{fig:dipcoating}c, a step gradient consists of
two half planes of different wetting properties. It is, therefore,
easy to fabricate via partly dip-coating of initially hydrophilic
substrates (silicon wafer, glass) with hydrophobic coatings as
schematically illustrated in \fref{fig:dipcoating}a. 

\blue{The substrate is dipped into the solution containing the methacrylate
    copolymer. The substrate is not dipped into a solution of the different
    components. The methacrylate copolymer contains a silane cross-linking
    component. This component leads not only to internal cross-linking of the polymer
    film but also to cross-linking of the substrate by the formation of
    siloxane bonds.}

For the hydrophobic coating, we use the block-co-polymer
Poly-(tert-butylmethacrylate-co-Zonyl$^{\mbox{\textregistered}}$-co-3-trimethoxysilylpropylmethacrylate)
6:2:2 provided by our co-worker R.\ Frenzel from the IPF Dresden
\cite{DuPont2002,Degischer2003}. The fabrication process is proven
to be reproducible and the surfaces show stable wetting properties
over long periods of time.

The substrates are characterized by contact angle measurements and
ellipsometry. Varying the block-copolymer concentration of the
toluene solution and the dip-coating speed, we are able to control
the thickness of the coating film in the range of 5nm to 24nm. The
wetting properties of the coated substrate are found to be
independent of the film thickness with an average advancing contact
angle of $105^\circ$ and an average receding contact angle of
$79^\circ$. The polymer films turned out to be stable after
exposition to water and solvents for 30 min at $65^\circ$C in the
ultrasonic bath. The polymer coatings have the advantage to form
stable homogeneous films on all substrates of interest such as
silicon wafers, glass, aluminum, silicon nitride and epoxy resins.

\hide{\begin{figure} \epsfig{file=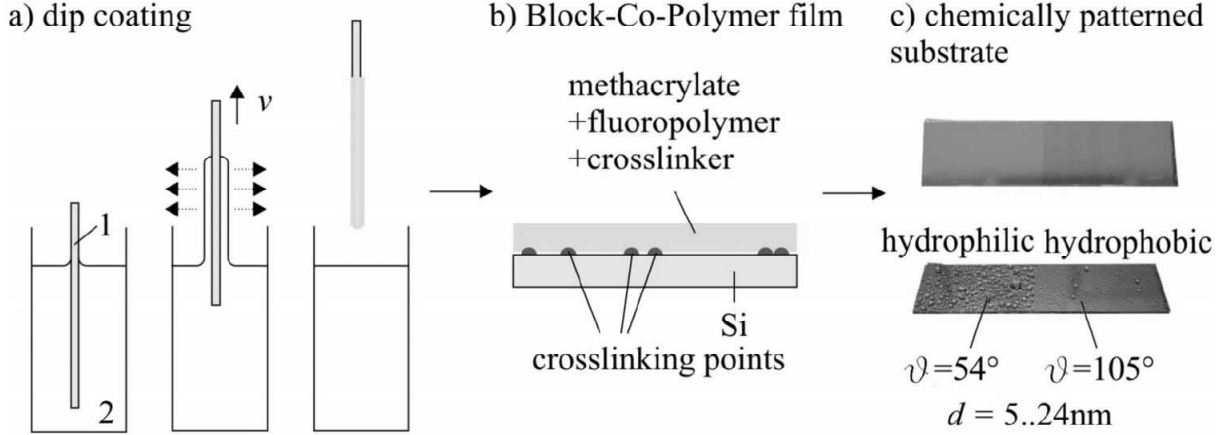,height=65mm,clip=}
\caption[]{a) Schematic illustration of the dip-coating process:
the substrate (1) is dipped into a polymer solution (2). By removing
the substrate at a speed $v$ liquid is dragged along with the
substrate. The solvent evaporates and the polymer concentration
increases until a polymer film forms. b) Schematic view of the thin
film structure: After annealing in water vapor, a
block-copolymerfilm builds up which consists of a methacrylate bulk,
hydrophobic fluoropolymer chains and former crosslinking agent. c)
Digital photograph of a silicon substrate, coated with the
block-copolymer film on the right half. In the lower picture, water
droplets have been sprayed onto the substrate to visualize the
wettability step gradient.} \label{fig:dipcoating}
\end{figure}}

\subsection{Theoretical method}
In parallel to the above mentioned experimental methods, we also
address basic questions relevant for the separation process
 theoretically. This includes simple
analytical estimates as well as lattice Boltzmann computer simulations.

In the past twenty years, the lattice Boltzmann method
\cite{McNamara1988,Higuera1989,Benzi1992,Qian1992} has proved itself
as a versatile theoretical tool for the study of a variety of fluid
dynamical phenomena such as the flow through porous media
\cite{Gunstensen1993}, flow of polymer solutions \cite{Ahlrichs1999}
as well as properties of suspensions of solid particles
\cite{Ladd2001}, to name just a few examples. Comprehensive
introductions to the general foundations of the lattice Boltzmann
method can be found in recent monographs
\cite{Rothman1997,Wolf-Gladrow2000,Succi2001} as well as in review
articles \cite{Chen1998,Raabe2004}.

While examples given above are mainly based on (extensions of) the LB method
for the so called ideal fluids (fluids with an ideal gas equation of state),
other LB approaches have been devised allowing the simulation of e.g.\
a liquid in coexistence with its vapor as well as
a binary fluid mixture \cite{Shan1993,Swift1995,Luo1998}.

In this work, we employ a free energy based lattice Boltzmann method
capable of adequately simulating both the static and the dynamic properties
of a liquid-vapor system. The method has first been
proposed by Swift and coworkers  who used the method for a study of
spinodal decomposition and domain growth \cite{Swift1995}. The
original version of the method suffered, however, from the lack of
Galilean invariance, a serious drawback, when hydrodynamic transport
becomes a relevant issue.

This problem was solved by Holdych and coworkers \cite{Holdych1998}
who proposed a modified expression for
the relation between the pressure tensor and the second moments of
the population densities in the lattice Boltzmann model. The model
was developed further by other coworkers and was applied for a study
of interesting issues such as the motion of
three phase contact line in two dimensions
\cite{Briant2002} and roughness effects on liquid-solid slippage
\cite{Dupuis2006}.

The reader interested in an elaborate discussion of the present lattice
Boltzmann approach is referred to many publications by Julia Yeomans and coworkers,
some examples of which were cited above \cite{Swift1995,Holdych1998,Briant2002,Dupuis2006}.
Here, we give a brief overview of the method only.

\blue{The starting point of the approach is a free energy functional of the form
\begin{equation}
\Psi = \int_{\Omega}  d^3r (\psib(\rho(\br)) + \frac{k}{2} |\bnabla \rho|^2) + \int_{S} dS \psis.
\label{eq:freeenergy}
\end{equation}
In \eref{eq:freeenergy}, $\psib(\rho)$ is the free energy per unit volume of a homogeneous 
system at density $\rho$ and $\psis$ the free energy per unit area associated with the
presence of the substrate. The first integral extends over the volume of the fluid 
(containing both the gas and the liquid phases) $\Omega$ and the second integral 
is evaluated over the surface of the substrate $S$.}

\blue{In the LB model adopted in the present simulations, the free energy density of 
the homogeneous phase $\psib$ is given by \cite{Briant2002}
\begin{equation}
\psib = \pc (\nurho+1)^2(\nurho^2-2\nurho+3-2\beta\nuT)
\label{eq:psib}
\end{equation}
and the resulting equation of state ($p=\rho \partial \psib / \partial \rho - \psib$) is} \cite{Leopoldes2003}
\begin{equation}
p = \pc (\nurho+1)^2(3\nurho^2-2\nurho+1-2\beta\nuT).
\label{eq:p}
\end{equation}
In Eqs.\ (\ref{eq:psib}) and (\ref{eq:p}) 
$\nurho=(\rho/\rhoc-1)$, $\nuT=(1-T/\Tc)$ and $\pc=1/8$, $\rhoc=7/2$ and $\Tc=4/7$ are the critical pressure, density and temperature respectively. The parameter $\beta$ is a constant allowing to tune e.g.\ the temperature dependence of equilibrium densities of the liquid and the vapor phases \cite{Briant2002}
\begin{equation}
\rho_\mathrm{L,G}=\rhoc\left(1\pm \sqrt{\beta\nuT} \right)
\label{eq:rhoLG}
\end{equation}
with $\beta=0.1$ in our simulations. As shown by Cahn and Hilliard
in their seminal work \cite{Cahn1958}, within a free energy based
model similar to that underlying the present LB approach, the
surface tension can be obtained from the knowledge of the density
profile as an integral over the density gradient across the
liquid-vapor interface (here assumed to be perpendicular to the
$z$-axis)
\begin{equation}
\sigma= \kappa \int \left(\frac{\partial \rho}{\partial z}\right)^2 dz.
\label{eq:surftension_A}
\end{equation}
Equivalently, given the input parameters of the present LB model,
$\sigma$ can also be obtained from the analytic expression \cite{Briant2002}
\begin{equation}
\sigma=\frac{4\rhoc}{3}\sqrt{2\kappa\pc}(\beta\nuT)^{3/2}.
\label{eq:surftension_B}
\end{equation}
A nice feature of Eqs.\ (\ref{eq:surftension_A}) and (\ref{eq:surftension_B}) is that they
provide an independent way to test the reliability of the simulation results.

In all the simulations to be reported below, we used $T=0.4$
corresponding to $\nuT=0.3$. This choice is below the critical temperature and gives rise to a liquid phase in equilibrium with its vapor with corresponding densities of $\rhoL \approx 4.1$
and $\rhoG \approx 2.9$. Unless otherwise stated, the LB relaxation time is set to $\tau=0.8$
(\eref{eq:collide}) and a value of $\kappa=0.004$ is used. \blue{The geometry of the
simulation is a rectangular three dimensional lattice. The topography
of the substrate is perfectly planar (no roughness) and parallel to the
$xy$-plane. It is placed both at the bottom of the box ($z=0$) and on the top
($z=L_z-1$). Periodic boundary conditions are applied along the $x$ and $y$
directions.} 

In order to save computation time, system size is varied depending on the specific situation with typical values around
$L_x\times L_y\times L_z = 100 \times 100 \times 100$.

\blue{As to the effect of a substrate on system properties,
minimizing the free energy functional $\Psi$ [\eref{eq:freeenergy}] 
subject to the boundary condition that $\psis=- \phi_1 \rhos $ \cite{Cahn1958,deGennes2002}, 
where $\phi_1$ is a constant and $\rhos$ the fluid density at the 
substrate, leads to the condition that the gradient of the density 
in the direction normal to the substrate ($\bs$) must
satisfy \cite{Leopoldes2003}
\begin{equation}
\kappa \bs \cdot \bnabla \rho = -\phi_1.
\label{eq:suba}
\end{equation}
}

\blue{Introducing the equilibrium contact angle $\theta$ and assuming
a planar substrate at $z=0$ parallel to the $xy$-plane 
($\bs \cdot \bnabla=\partial / \partial z$),} it can be shown that
\cite{Leopoldes2003}
\begin{equation}
\frac{\partial \rho}{\partial z} = -2 \beta \nuT \sqrt{\frac{2\pc}{\kappa}} \mathrm{sign}(\theta-\frac{\pi}{2})
\sqrt{\cos(\frac{\alpha}{3})[1-\cos(\frac{\alpha}{3})]}.
\label{eq:sub}
\end{equation}
The angle $\alpha$ is determined by the input contact angle
$\theta$ via $\cos(\alpha)=\sin^2(\theta)$. \blue{Note that, in contrast to
\eref{eq:suba}, the dependence of the density gradient on $\phi_1$ is no
longer explicit. Rather, it affects $\partial \rho/\partial z$ at 
the substrate via the equilibrium contact angle $\theta$. This is a nice
feature allowing the use of $\theta$ as input parameter of the simulation.}

Turning our attention to the simulation technique, we now address the 
central quantity within a lattice Boltzmann scheme, the so called distribution
function (or population density) $f_i$. Imagine a volume of fluid
around a point $\br$ at time $t$ and divide the fluid within this volume
to a finite number of portions (parcels).
Roughly speaking, $f_i(\br,t)$ would then denote the fluid
fraction (parcel) moving with a velocity $\bci$.
Obviously, once the populations $f_i(\br , t)$ are known, one obtains the fluid density
$\rho$ and  velocity $\bu$ via
\begin{eqnarray}
\rho(\br,t)&=&\sum_{i=0}^{b} f_i(\br,t)
\label{eq:rho}\\
\rho(\br,t)\bu(\br,t)&=& \sum_{i=1}^{b} f_i(\br,t)\bci,
\label{eq:u}
\end{eqnarray}
where we already assumed a regular lattice, where each site is linked to $b$ neighboring sites.
Here, the index $i$ is used to enumerate various links $\bci$ along which
a lattice node is connected to its neighbors.
The reader may have noticed that we
do not distinguish between velocity and link vectors. This is a result
of the underlying LB scheme, where a fluid portion moving along the link number $i$
travels the whole length of the link during one single time step
(this gives rise to a higher speed along diagonal directions compared
to the main coordinate directions; see also below).

In the present LB model, we use a regular cubic lattice, where a given node is
connected to its neighbor nodes along the 6 coordinate directions
$\bci \in \{(\pm 1, 0, 0), (0,\pm 1, 0), (0, 0, \pm 1)\}$ ($i=1,...,6$)
as well as along the 8 diagonal directions $\bci \in \{(\pm 1, \pm 1, \pm 1)\}$ ($i=7,...,14$).
The index $i=0$ is reserved for the null vector, $\bc_0=\vec{0}$, corresponding to the so called rest particles. This defines to so called three dimensional 15 velocity (D3Q15) LB model.

Within a lattice Boltzmann method, one basically iterates two simple steps
generally referred to as (1) relaxation (collision) and (2) free propagation (streaming),
\begin{eqnarray}
&&f_i'(\br,t) = f_i(\br,t) - \frac{1}{\tau}[f_i(\br,t)-\fieq(\br,t)]
\label{eq:collide}\\
&&f_i(\br+dt\bci, t+dt) = f_i'(\br,t),
\label{eq:stream}
\end{eqnarray}
where we introduced $f_i'$ in order to formally separate the relaxation and
streaming steps. In \eref{eq:collide}, the quantity $\tau$ is the so called relaxation time
which determines the fluid's kinematic viscosity, $\nu=\eta/\rho$ ($\eta$=viscosity, $\rho$=fluid density).
For the present three dimensional 15 velocity model, one finds \cite{Briant2002}
$\nu = (\tau-0.5)dx^2/(3dt)$, where $dx$ is the distance between two neighboring
nodes connected along the main coordinate directions and $dt$ is the time step.

Physical properties of the system enter the LB iteration scheme via the
quantity $\fieq$ (\eref{eq:collide}). Obviously,
the system is 'pushed' towards $\fieq$ with a rate $1/\tau$.
The population density $\fieq$ is, therefore, referred to as 'equilibrium
distribution'. \blue{It is noteworthy that the term 'equilibrium'
does not refer to a global thermal equilibrium, where no flow exists.
Rather, it describes the \emph{local} velocity distribution in a portion 
of fluid moving at a velocity $\bu(\br)$.} Within the present LB model, 
one expands $\fieq$ in powers of the fluid velocity $\bu$ up to 
the second order  \cite{Briant2002}:
\begin{equation}
 \fieq=A_{\sigma} + B_{\sigma} \bu \cdot \bci + C_{\sigma} u^2  + D_{\sigma} (\bu \cdot \bci)^2 + \bci \cdot \tensor{G}_{\sigma} \cdot \bci
\label{eq:fieq}
\end{equation}
for $i>0$. The population of the rest particles 
\blue{then results from the constraint that the relaxation 
(collision) process does not change the fluid density. This leads
to $ \sum^{b}_{i=0} \fieq=\sum^{b}_{i=0} f_i = \rho$ ($b=14$ for the D3Q15 LB model), 
whereby leading to}
\begin{equation}
 f_0^{\mathrm{eq}}=\rho-\sum^{b}_{i=1} \fieq.
\label{eq:fieq0}
\end{equation}
\blue{Note that, even though the fluid density is unchanged during collision, it
may undergo a variation through the streaming step.}
The use of the index $\sigma$ instead of $i$ in \eref{eq:fieq}
is motivated by the fact that, due to symmetry requirements, we expect
exactly the same coefficients for all the links along principal directions, $i=1,...,6$.
In other words, $A_1=A_2=...=A_6$, $B_1=B_2=...=B_6$, etc. The value $\sigma=1$ is thus used
for $i \in \{1,...,6\}$. Similarly, $\sigma=2$ corresponds to all the links along diagonal directions,
reflecting $A_7=A_8=...=A_{14}$, $B_7=B_8=...=B_{14}$, etc. A possible choice of the coefficients
$A_{\sigma},\; B_{\sigma},\; C_{\sigma},\; D_{\sigma}$ and the tensor $\tensor{G}_{\sigma}$
is  (no summation over repeated indices) \cite{Briant2002}

\begin{eqnarray}
A_{\sigma} &=& \frac{w_{\sigma}}{c^2} \left\{ p - \frac{\kappa}{2} ||\bnabla \rho||^2 -\kappa \rho \Delta \rho + \nu \bu \cdot \bnabla \rho \right\} \label{eq:A}\\
B_{\sigma}&=&\frac{w_{\sigma}\rho}{c^2}, \;\; C_{\sigma}=-\frac{w_{\sigma}\rho}{2c^2},\;\; D_{\sigma}=\frac{3w_{\sigma}\rho}{2c^4}\label{eq:BCD}\\
G_{1\gamma\gamma}&=&\frac{1}{2c^4}\left\{
\kappa (\partial_{\gamma} \rho)^2 + 2\nu u_{\gamma} \partial_{\gamma} \rho
\right \} \label{eq:G1}\\
G_{2\gamma\gamma}&=&0 \label{eq:G2A}\\
G_{2\gamma\delta}&=&\frac{1}{16c^4}\left\{
\kappa (\partial_{\gamma}\rho) (\partial_{\delta}\rho) + \nu (u_{\gamma} \partial_{\delta} \rho + u_{\delta} \partial_{\gamma} \rho) \right \}\nonumber\\ && (\gamma\neq \delta).
\label{eq:G2B}
\end{eqnarray}
where Greek letters label Cartesian coordinates $x_{\gamma}, x_{\delta} \in \{ x,y,z \}$ and
$\partial_{\gamma,\delta} \equiv \partial /  \partial x_{\gamma,\delta}$. Note that the tensor $\tensor G_1$ couples to velocities $\bci$ parallel to the
coordinates axis ($i=1,...,6$) only. The non-diagonal components of $\tensor G_1$
are, therefore, of no interest here.

In  Eqs.\ (\ref{eq:A})-(\ref{eq:G2B}), $w_1=1/3$ and $w_2=1/24$ are constant weights and
$c=\delta x / \delta t$ is the velocity along a principal direction
(note that the velocity along a diagonal line is $\sqrt{3}c$).
$\kappa$ is a parameter which tunes the width of the liquid-vapor
interface and the related surface tension (see also below). In addition to the
presence of spatial derivatives of fluid density, non ideal effects are also
accounted for in the expression for the pressure $p$ [\eref{eq:p}]
which enters the equilibrium distribution
$\fieq$ via the coefficient $A$ [see Eqs.\ (\ref{eq:fieq}) and (\ref{eq:A})].

\section{Results and discussion}
In order to obtain  a first, order of magnitude, estimate of the
driving forces present in the separation process, we give in the
next section simple analytical estimates of these forces for the two
interesting cases of open substrate and closed planar geometry (slit).
Results of experiments as well as lattice Boltzmann simulations will
then be presented in the subsequent sections.

\subsection{Driving forces of the separation process}
\label{subsection:drivingforce} In a typical separation experiment,
a binary emulsion enters the gradient zone in a direction
perpendicular to the direction of wetting gradient (see e.g.\ the
lower image in \fref{fig:sep_exp}b). In the absence of a
wetting gradient, each component of the emulsion would have equal
probabilities of selecting either the right or the left arm of the
channel. The role of the wetting gradient is to induce a
preferential deflection of at least one of the fluid
components.

Note that, as will be shown below (Figs.\
\ref{fig:sep_exp2} and \ref{fig:sep_exp3}), selective wetting of one
of the fluid components is sufficient for a successful separation.
To keep the analysis as simple as possible, it is, therefore,
reasonable to first focus on the effects of a
step-wise wetting gradient on the motion of a single fluid droplet.

Suppose that a fluid droplet is in contact with both hydrophilic and
hydrophobic parts of a substrate with a step-like wetting gradient. Due to
different contact angles on the both sides of the wetting gradient, there will
be a net force on the droplet pushing it towards the more hydrophilic
(or, equivalently, less hydrophobic) part of the substrate.
This force is proportional to the derivative of the underlying 
thermodynamic potential (the sum of all the surface free energies involved 
in the problem) with respect to the equilibrium contact angle.

On the other hand, it is not unreasonable to assume that the 
droplet can locally adapt its contact angle to the local equilibrium 
value within a time short compared to the translation of the droplet 
over the chemical step. In this case, the droplet shape will be a 
kind of compromise (interpolation) between two spherical caps, 
corresponding to the contact angles on both sides of the wettability step.
Focusing on the proximity of the substrate, the droplet shape far from
the chemical step will be close to that of a spherical cap with a 
contact angle-dependent radius of curvature. As a consequence, a gradient in the Laplace pressure 
(see below) will form \emph{inside} the droplet, trying to push it towards the more 
hydrophilic side of the substrate (see e.g.\ \fref{fig:lbm_spreading_on_step_grad}).

It is, therefore, instructive to start with simple analytical estimates of the Laplace pressure
and its dependence on the contact angle and other relevant parameters.

\begin{figure}
\epsfig{file=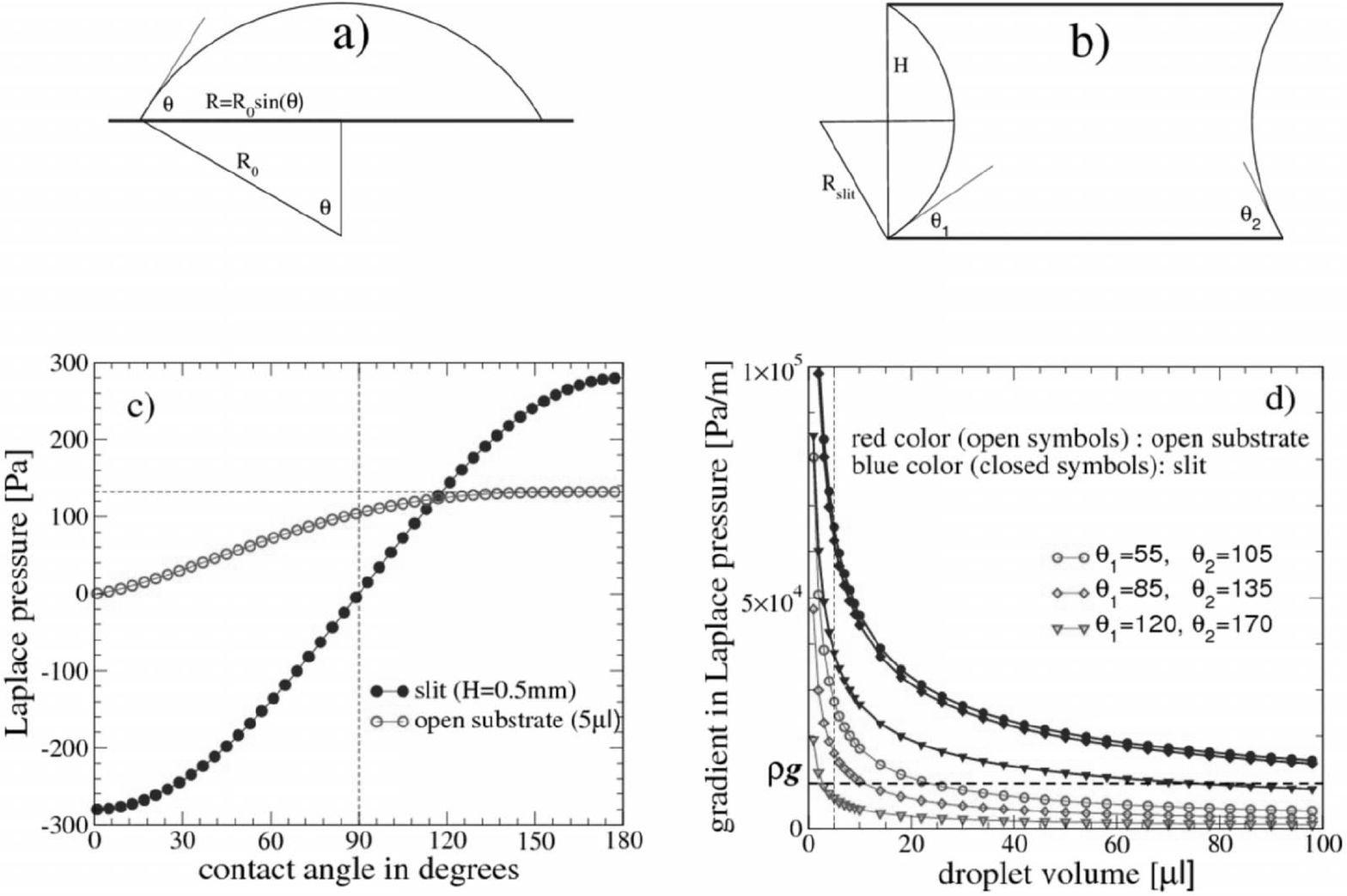,height=110mm,clip=}
\vspace*{-10mm}
\caption[]{a) and b) Schematic views of a droplet on a chemically homogeneous
open substrate [a)] and a liquid film confined in a planar gap
with a wetting gradient [b)]. c) The
Laplace pressure versus contact angle both for a 5$\mu$l spherical
cap of water placed on an open substrate  and within a
liquid film of water confined in a planar gap of height $H=0.5$mm
as indicated. A  vertical dashed line marks $\theta=90^\circ$.
A horizontal dashed line marks the limiting value of the Laplace
pressure in the case of spherical cap for $\theta \to 180^\circ$.
It serves to underline the presence of a plateau in this limit
on open substrate. d) The gradient of Laplace pressure versus droplet volume
within a spherical cap [Eqs.\ \ref{eq:pLcap} and \ref{eq:fxcapa}] and in a planar slit
[Eqs.\ \ref{eq:pLslit} and \ref{eq:fxslita}] for a contact angle difference of
$d\theta=50^\circ$ as indicated. For comparison, a horizontal
dashed line marks the gravitational force \blue{density}. A vertical dashed line marks
$\Omega=5\mu$l, for which the $\theta$-dependence of $\pL$ is shown in
the panel c).}
\label{fig:pL}
\end{figure}

Let us first consider the case of a droplet forming a spherical cap on a 
chemically \emph{homogeneous} open substrate (the panel a) in \fref{fig:pL}).
For a spherical cap of volume $\Omega$ and static contact angle $\theta$, the
pressure inside the droplet is $p=p_0+\pL(\theta, \Omega)$, where $p_0$ is the
pressure outside the droplet, and $\pL=2\sigma/R_0$ the so called
Laplace pressure arising from the curvature $R_0$ of the liquid-vapor interface 
($\sigma=$surface tension. The factor of 2 originates from the presence of two 
equal radii of curvature).

Note that, in the above, we express the pressure inside the droplet
in terms of the radius of curvature and not the radius of contact area.
The latter being given by $R=R_0\sin(\theta)$, one recovers the (perhaps more familiar)
expression $p=p_0+2\sigma\sin(\theta)/R$.

The radius of curvature of a spherical cap is given by
\begin{equation}
 R_0=\left( \frac{\Omega}{\pi(2/3+\cos(\theta)^3/3-\cos(\theta))}\right)^{1/3}
\label{eq:R0}
\end{equation}
as can be obtained via integrating a portion of the sphere for polar
angles in the interval [0, $\theta$]. Inserting this information into the
expression for the Laplace pressure within a spherical cap, we obtain
\begin{equation}
\pL=\frac{2\sigma}{R_0}=2\sigma\left(\frac{\pi(2/3+\cos(\theta)^3/3-\cos(\theta))}{\Omega}\right)^{1/3}.
\label{eq:pLcap}
\end{equation}

The panel c) in \fref{fig:pL} shows, for the case of water, the Laplace pressure
versus contact angle for a
spherical cap of $\Omega=5\mu$l volume (circles). For contact angles close
or below $\theta=90^\circ$, the Laplace pressure varies rather fast
but then approaches a quasi-plateau as the hydrophobic zone is
reached. The horizontal dashed line serves to highlight this
limiting behavior.

Note that it is the \emph{difference} in Laplace pressure,
$\delta p  = \pL(\theta_2)-\pL(\theta_1)$, which
is the origin of a driving force. More precisely, for a given variation in the
contact angle, the driving force \blue{per unit volume (force density)}, 
$\bfdriv$, is proportional to the \emph{gradient} of the Laplace pressure:
\begin{equation}
\bfdriv = -\bnabla \pL = -\frac{\partial \pL}{\partial \theta} \bnabla \theta.
\label{eq:fdriv}
\end{equation}
Since ${\partial \pL} / {\partial \theta}$ on open substrates
approaches zero for large contact angles, increasing the contact angle of the
more hydrophobic part of the substrate (i.e.\
making it still more hydrophobic) does not necessarily lead to a
significantly higher driving force \blue{density}.

This idea is nicely born out in panel d) of \fref{fig:pL}, where we show
(for the case of water) the gradient of the Laplace pressure
versus droplet volume (see also \eref{eq:fxcap} below)
for the \emph{same difference} in contact
angle $d\theta=\theta_2-\theta_1=50^\circ$ while shifting both $\theta_1$
and $\theta_2$ towards progressively higher values. Obviously, at constant $d\theta$,
the higher $\theta_1$ the lower  the wettability gradient induced driving
force \blue{density}.

Using Eqs.\ (\ref{eq:pLcap}) and (\ref{eq:fdriv}), we can also estimate
the dependence of the wettability induced driving force \blue{density} on the volume
of a spherical cap
\begin{eqnarray}
\fxdriv &\simeq & -\frac{\pL(\theta_2)-\pL(\theta_1)}{l_x}
\label{eq:fxcapa}\\
&\propto& \frac{1}{R_0^2} \propto \frac{1}{\Omega^{2/3}}\; \mbox{(open substrate)}.
\label{eq:fxcap}
\end{eqnarray}
In deriving the scaling relation \ref{eq:fxcap}, it is assumed that
the lateral extension of the droplet (the length $l_x$ over which
the difference in Laplace pressure builds up) is equal to the sum of
the radii of contact, $R_1$ and $R_2$, corresponding to contact
angles $\theta_1$ and $\theta_2$, respectively:
\begin{equation}
l_x \simeq (R_1+R_2) = R_0(\theta_1)\sin(\theta_1)+R_0(\theta_2)\sin(\theta_2).
\label{eq:lx}
\end{equation}
In contrast to open surfaces, the radius of curvature of the
liquid/air front for a liquid film inside a channel made of parallel
plates (slit geometry) does not depend on the droplet volume, provided
that the liquid volume is large enough to fill the gap vertically
and reach a lateral extension larger than the gap height. Moreover, for
channel heights smaller than the capillary length, the gravitational
effects on the shape of the liquid/air interface may be
neglected \cite{deGennes2002}.

Under these
circumstances, the radius of curvature of the
liquid/air front in a planar slit is given by
$R_0=H/(2\cos(\theta))$. As a consequence, the
Laplace pressure within a liquid film enclosed in a
planar channel is obtained as
\begin{equation}
\pL=-\frac{2\sigma \cos(\theta)}{H}\;\;\;\;\;(\mbox{slit geometry}).
\label{eq:pLslit}
\end{equation}

To make a parallel to \eref{eq:fxcap} for the case of slit geometry,
we estimate the dependence of the resulting
wettability induced driving force \blue{density} on the volume of fluid enclosed in a planar slit.
Assuming that the fluid takes the shape of a circular disk, the lateral extension
of a droplet of volume $\Omega$ within a narrow slit of height $H$ can be estimated
from $\pi r^2 H=\Omega$. This gives $l_x^\mathrm{slit} \simeq 2 r = \sqrt{4\Omega/(\pi H)}$ and hence
\begin{eqnarray}
\fxdriv &\simeq& - \frac{\pL(\theta_2)-\pL(\theta_1)}{\sqrt{4\Omega/(\pi H)}}
\label{eq:fxslita}\\
&\propto& \frac{1}{(H\Omega)^{1/2}}\; (\mbox{slit}),
\label{eq:fxslit}
\end{eqnarray}
where \eref{eq:pLslit} for $\delta \pL$ is used. \blue{
Note that the scaling laws \erefs{eq:fxslit}{eq:fxcap} can also be obtained via an 
estimate of the gradient of surface free energies involved in the process.}

Let us examine whether we are allowed to use the above estimates in the case of
our experiments. For this purpose we first estimate the capillary
length in order to see whether neglecting gravity is justified. For
water, using the values of the surface tension
$\sigma=0.07\mathrm{N/m}$, mass density $\rho=10^3\mathrm{kg/m}^3$
and gravitational acceleration $g=9.8\mathrm{m/s}^2$, the capillary
length can be estimated as $\lcap=\sqrt{\sigma/g\rho} \approx
2.67$mm. Since the height of the channel used in our experiments is
$H=0.5$mm and thus significantly smaller than $\lcap$ (see
\fref{fig:sep_exp2}), gravity does not play a major role here.

As to the independence of the radius of curvature from the
liquid volume, the size of the smallest fluid droplet leaving our
emulsion generator is $\Omega\approx 2.9\mu$l (see \fref{fig:emulsion-generation}).
Once entered the planar slit, such a droplet spreads horizontally
taking approximately the shape of a circular disk
(cylinder) with a radius, $r$, given by $\pi r^2 H \approx \Omega$.
Setting the numbers $\Omega=2.9\mathrm{mm}^3$ and $H=0.5$mm,
one obtains $r \approx 1.36$mm which is roughly three times the gap height.
This justifies the assumption that the shape of the liquid-vapor
interface is independent of the droplet volume.

Thus, even for the smallest droplet in our
experiments, the Laplace pressure within the slit and
the corresponding wettability gradient induced driving forces
can be estimated via Eqs.\ (\ref{eq:pLslit}) and (\ref{eq:fxslit}).

The dependence of the Laplace pressure upon the contact angle
for the case of a closed planar geometry is shown in the
 panel c) of \fref{fig:pL} for a channel of height
$H=0.5$mm as used in our experiments. Obviously, for
the range of parameters studied in our experiments, the slit
geometry allows the formation of significantly higher gradients in
Laplace pressure than would be possible on an open substrate.

Furthermore and in sharp contrast to open geometry, this applies to
contact angle variations both in the hydrophilic and in the
hydrophobic regime: As shown in the panel d) of  \fref{fig:pL},
for a difference of $\theta_2-\theta_1=50^\circ$,
the curves for $\theta_1=55^\circ$ and $\theta_1=85^\circ$ are
practically identical in the case of a slit geometry while
on open substrate the gradient of the Laplace pressure drops significantly when
going from $\theta_1=55^\circ$ to $\theta_1=85^\circ$.

A significant decrease of $\nabla \pL$ is, however, observed also in
the case of a slit when choosing $\theta_1=120^\circ$ and
$\theta_2=170^\circ$. This is not surprising since the Laplace
pressure as a function of the contact angle gradually flattens for
large contact angles. This point is nicely seen in the panel c) of
\fref{fig:pL}.

While neglecting many complications occurring in real experiments,
this simple estimate yields at least a qualitative
understanding of higher separation efficiency observed
in confined channels as compared to open substrates.

\subsection{Separation experiments on step gradients}
A variety of separation experiments on surfaces with step gradients
have been performed. All these experiments show efficient separation
processes with high reproducibility \cite{PagraDiss}. Additionally,
it is possible to vary the flow velocity of the emulsion over
a wide range. The efficiency of step gradients is first tested on
open substrates using different lateral geometries and wettability
contrasts between the hydrophilic and hydrophobic parts of the
substrate (\fref{fig:sep_exp}).

A demixing of the water-toluene emulsion is observed in all the
experiments performed on open substrates. Furthermore, these
experiments suggest that the selectivity of water with respect
to the both parts of the substrate mainly drives the separation. We
will come back to this important point below.

\hide{\begin{figure} 
\epsfig{file=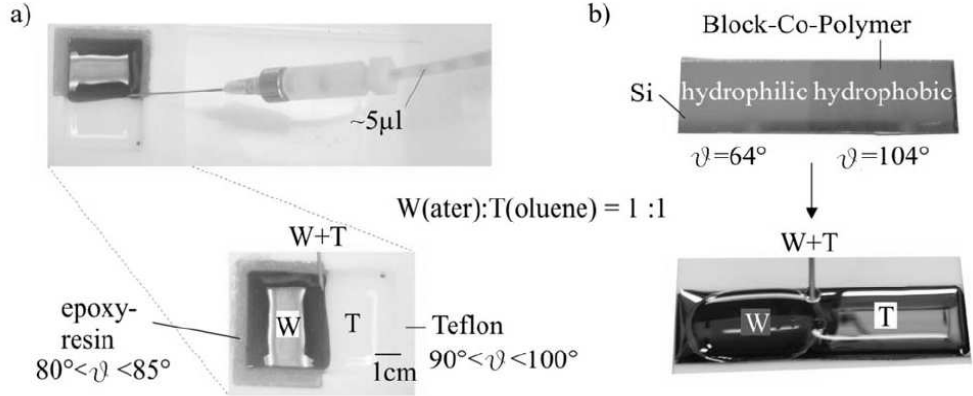,height=60mm,clip=}
\caption[]{
a) Separation of a water-toluene emulsion in a
microfluidic cell. The cell is milled into a Teflon block
(hydrophobic) with a cavity filled with epoxy resin (hydrophilic). A
toluene/water mixture serves as emulsion, where blue ink dissolved
into the water is used to enhance the contrast. While toluene wets
both surfaces roughly in the same way, water exhibits a strong
tendency to selective wetting. (b) Top: A silicon substrate
partially coated by a fluorosilane (right half). The uncoated part
is strongly hydrophilic (contact angle of 64$^\circ$) while the
coated part is hydrophobic (104$^\circ$). Bottom: A separation
experiment performed on this substrate.}
\label{fig:sep_exp}
\end{figure}}

\blue{With respect to applications, experiments in confined geometries 
are of major importance because confinement prevents fast liquid 
evaporation that would occur in open  miniaturized systems.}
Therefore, in a further step, we turned our attention to 
separation experiments in confined geometries, i.e. in closed cells.

\blue{Separation cells with widths (horizontal extensions) of 1mm, 2mm, 3mm, 4mm, 8mm, 
14mm and 20mm were used. However, for the separation cells 
with a width less than 20mm, a stable reproducible separation 
process could not be established. Qualitatively, we observe a decrease 
of separation efficiency upon a reduction of channel width. 
This suggests that adhesive forces due to surface energy hysteresis oppose
the driving forces induced by the step gradient of wettability. A detailed
analysis of this issue can be found in \cite{PagraDiss}.}

\blue{In the case of the largest separation cell investigated (width=20mm),
the separation efficiency is 100\% for flow rates up to 2 ml/min and 
no 'mislead' droplets are observed. For flow rates between 5 ml/min 
and 10 ml/min single droplets of the 'wrong' component are observed 
on the hydrophobic and the hydrophilic side. The estimated separation 
efficiency in this case remains well above 90\%. 
Currently, a separation cell with integrated reservoirs for each outlet 
is being developed which will enable us to specify the separation efficiency 
more precisely. However, this modification is not trivial because 
the attachment of additional fluidic components changes the boundary 
conditions of the whole problem.}

Figure
\ref{fig:sep_exp2} shows a digital photograph of the separation
chamber. It consists of an aluminum block (1) into which a planar
channel has been milled (with a tiny hole in the middle
that serves as the inlet). The aluminum
channel is covered by a glass substrate (2) which is glued onto the
aluminum block by epoxy resin (3). The aluminum and the glass part
are both half-coated with the block-copolymer film.
The dimensions of the so constructed channel
are $L=75$mm (length), $W$=20mm (width) and $H=0.5$mm (height).

At early stages of the separation (a), water (W) exclusively wets
the hydrophilic side. Interestingly, the non-polar component of the
mixture (here hexane (H)) also shows a tendency to flow along with
water and wets the hydrophilic part of the channel (see the
liquid/air front in \fref{fig:sep_exp2}a. However, the situation
changes as soon as water reaches both sidewalls. At this point, the
water front formed across the channel hinders the motion of the
non-polar component towards the hydrophilic side. The latter thus
flows towards the hydrophobic side and the separation becomes
perfect (\fref{fig:sep_exp2}b).

\hide{\begin{figure} \epsfig{file=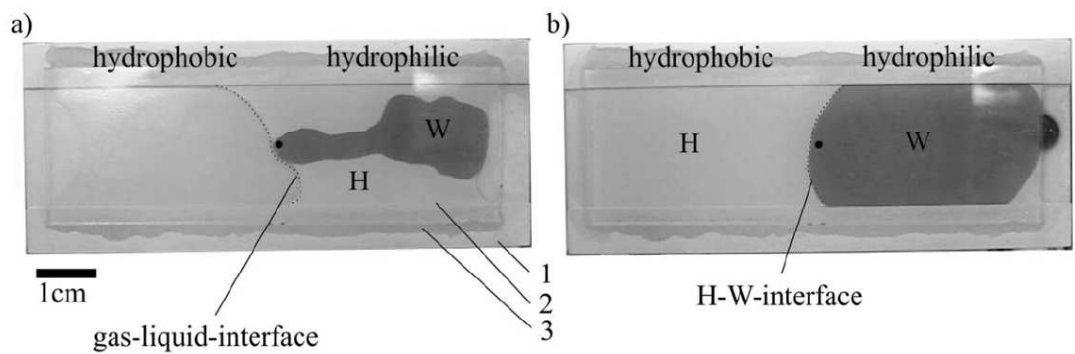,height=50mm,clip=}
\caption[]{
W=Water, H=Hexan, 1=Aluminum channel 2=Glass cover,
3=Clay (Epoxid). The left halves of both the Aluminum channel and
the glass cover are covered with a hydrophobic coating
(fluoropolymer). (a) Initial stage of separation. (b) Final state.
The channel has a length of $L=75$mm, a width of $W$=20mm and
a vertical dimension of $H=0.5$mm.
}
\label{fig:sep_exp2}
\end{figure}}

\subsubsection{The effect of a secondary flow on separation efficiency}
Motivated by the above discussed observation, that the separation
becomes practically perfect as soon as the water component reaches
both side walls, we introduce an additional water source in order to
enhance the separation efficiency. Figure \ref{fig:sep_exp3} shows a
digital photograph of a separation chamber designed for this
purpose. The chamber has two inlets. Through the one inlet the
hexane-water-mixture is introduced, whereas through the other inlet
only water is injected. For visualization purpose, the water
component in the mixture is colored with ink. As soon as the two
water components come into contact, they form a continuous front
across the channel pushing hexane to the right side of the channel.

At later stages of flow separation, a sharp interface builds up
between the hexane and the aqueous phase. The straight interface
between the two water components (indicated as W and W2) reflects
the laminar nature of the flow, a characteristic feature of most
(but not all, see e.g.\ \cite{Varnik2006a,Varnik2007a,Varnik2007b})
microflows. Due to the absence of chaotic behavior, the two water
components mix via diffusion only, a process slow compared to the
time scale of the separation experiment.

\hide{\begin{figure}
\epsfig{file=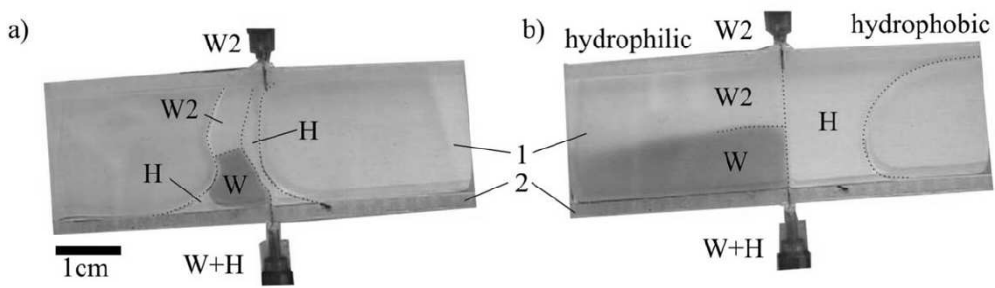,height=50mm,clip=}
\caption[]{
W=Water with ink, H=Hexane, W2=Water as secondary
flow, 1=Glass cover, 2=Sidewalls made of epoxy resin. Sidewalls,
glass cover and the bottom substrate are half coated with the
hydrophobic block-copolymer film. (a) separation starts as soon as
both water components W and W2 meet. (b) nearly perfect separation
result.} \label{fig:sep_exp3}
\end{figure}}

\subsection{Lattice Boltzmann simulations}
\label{subsection:simulation}
\subsubsection{Simulation of wetting gradient driven fluid motion}
In section \ref{subsection:drivingforce} we addressed, via simple
analytic arguments, the driving forces resulting from a change in
wetting properties of the substrate. These estimates are, however,
based on the assumption of perfect spherical caps on each side of
the step gradient with respective contact angles. Obviously, a
single droplet covering both sides of a step gradient can not
satisfy this condition. Rather, it will try to take a shape
consistent with variable wettability of the substrate. This shape
will probably resemble to a portion of a spherical cap at each end
of the droplet (away from the interface between the hydrophilic and
hydrophobic parts of the substrate) with a transition between the
two different 'spherical caps' in the step gradient zone.
Furthermore, dynamic effects, such as effects of dissipation loss on
the force balance, are fully absent in the above analysis.

This underlines the need for a full theoretical treatment of the
problem including the whole complexity of the droplet shape
variation across the transition zone as well as the internal fluid
dynamics. The lattice Boltzmann simulations to be presented below
provide such an approach.

\subsubsection{Droplet movement on step-gradients}
As already mentioned, a wetting gradient breaks the symmetry of the
problem introducing a preferential deflection of a water droplet
towards the more hydrophilic side of the channel.

In order to elucidate this issue, we first performed a series of
lattice Boltzmann computer simulations of the droplet motion on a
step gradient in the \emph{absence of external forces}. For this
purpose, a spherical fluid droplet is deposited  at time $t=0$
on the top of a step-like wetting gradient in a way that it slightly
touches the line separating the zones of different wettability. The
initial velocity of the droplet is zero, so that all dynamic effects
originate from the tendency of the liquid to wet the substrate.
Furthermore, both the hydrophilic and the hydrophobic halves of the
substrate are chosen to be perfectly flat thus ensuring the absence
of hysteresis effects. They differ only in their static contact angles with
respect to the fluid.

A typical result of these simulations is illustrated in
\fref{fig:lbm_spreading_on_step_grad} for a choice of static contact angles
$\theta_1=75^\circ$ (right half) and $\theta_2=105^\circ$ (left
half). Not unexpectedly, the droplet preferentially wets the
hydrophilic part of the substrate as illustrated by a sequence of
images in \fref{fig:lbm_spreading_on_step_grad}. This leads to
an asymmetric shape during the spreading process and a net
horizontal momentum of the droplet's center of mass towards the
hydrophilic part. This horizontal motion stops as soon as the
droplet fully leaves the hydrophobic part of the substrate
indicating that inertial effects are negligible in the case studied
here.

At this stage, it is instructive to estimate the average time $\bar{t}=l_x/(2u)$
($u$ is the velocity scale and $l_x$ the lateral extension of the droplet
on the substrate, see \eref{eq:lx}) for the motion of the droplet
over the step gradient. For this purpose,
we recall that inertial effects are negligible in our simulations.
Equating the viscous dissipation  $\eta \Delta u$ to the gradient
of the Laplace pressure $\nabla \pL$, we can then obtain an estimate
of the velocity scale $u$ if we realize that both $u$ and the Laplace pressure $\pL$
vary over the same length scale, $l_x$. Thus, $\Delta u \simeq u/l_x^2$
and $\nabla \pL \simeq [\pL(\theta_2)-\pL(\theta_1)]/l_x$ (recall that
the latter relation has also been used in deriving the scaling
relation \ref{eq:fxcap}). Along with \eref{eq:pLcap}, this yields
\begin{equation}
u \simeq  \ucap l_x [ 1/R_0(\theta_2) - 1/R_0(\theta_1) ],
\label{eq:u}
\end{equation}
where we used the definition of the capillary velocity $\ucap=\sigma/\eta$.
Introducing a closely related capillarity time, $\tcap=l_x/(2\ucap)$ allows to write
\begin{equation}
\bar{t} \simeq \tcap \frac{1}{l_x[1/R_0(\theta_2) - 1/R_0(\theta_1)]}.
\label{eq:tav}
\end{equation}
Let us estimate the characteristic velocity, $\ucap=\sigma/\eta$, and the
characteristic time, $\tcap$ for the case shown in
\fref{fig:lbm_spreading_on_step_grad}. For this purpose, we first estimate the
surface tension. Within the present lattice Boltzmann model and
for the specific choice of the parameters used in these simulations
(here: $\kappa=0.002$) we obtain via \eref{eq:surftension_B} $\sigma \approx 5.42\times 10^{-4}$
(note that all simulated quantities are measured in LB units. In particular, the
lattice spacing, $\delta x \equiv 1$, and the LB time unit, $\delta t \equiv 1$).

It is worth noting that an estimate of the surface tension via the integral over
the density gradient across the liquid-vapor interface [\eref{eq:surftension_A}] leads
to $\sigma \approx 5.47 \times 10^{-4}$ (LB units) which is identical to the exact value
within an error of $1\%$.

As to the shear viscosity, it is estimated by inserting
$\tau=0.8$ and $\rhoL \approx 4.1$ into the relation
$\eta = \rho \nu = \rho (\tau-0.5)/3$ which gives $\eta \approx 0.41$ in LB units.

Using the above results on the surface tension and the droplet viscosity, the capillary
velocity is estimated via $\ucap=\sigma / \eta \approx 1.32 \times 10^{-3}$ (LB units)
leading to a capillary time of $\tcap \approx 2.2 \times 10^4$, where we used
$l_x\approx 58$ as obtained via \eref{eq:lx} for a droplet of volume
$\Omega=4\pi r_0^3/3$ with $r_0=30$ LB units
(see \fref{fig:lbm_spreading_on_step_grad}) on a step gradient substrate with
contact angles $\theta_1=75^\circ$ and $\theta_2=105^\circ$ \label{page:tcap}.
Finally, using this value of $l_x$ along with \eref{eq:R0} we are able to
evaluate \eref{eq:tav} in order to obtain $\bar{t}\approx 2.5 \tcap$.

As simulation results presented in \fref{fig:lbm_spreading_on_step_grad}
clearly demonstrate, this rough estimate does indeed yield the
correct time scale for the separation process over a step-wise
wettability gradient. Thus, in addition to the fact that LB simulation results
shown in \fref{fig:lbm_spreading_on_step_grad} are in qualitative
agreement with experimental observations of a preferential
liquid motion induced by a step-wise change in
the wetting properties of the substrate
(Figs.\ \ref{fig:sep_exp}-\ref{fig:sep_exp3}),  they also
provide a means to address issues related to the separation
process in a more quantitative way.

\hide{\begin{figure}
\epsfig{file=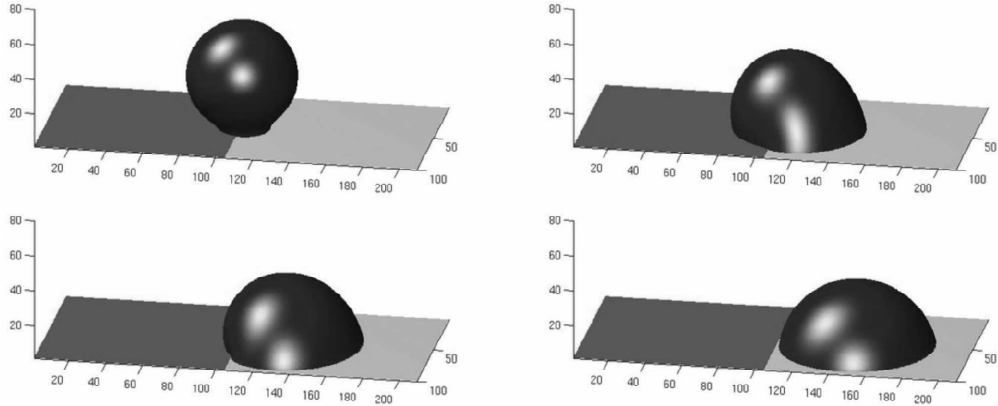,height=60mm,clip=}
\caption[]{Lattice Boltzmann simulation results on
the spreading of a droplet  on a step gradient. The left and right
parts of the substrate are
  characterized by static contact angles of $105^\circ$ (dark; red online) and
  $75^\circ$ (light (grey online)) respectively. At time  $t=0$, a droplet of radius $r_0=30$ LB units
is positioned on the top of a step gradient slightly touching
  the substrate (left). As time proceeds, the droplet develops a lateral motion
  towards the more hydrophilic side. The last image on the right corresponds to the final 
(equilibrium) state with zero velocity. From left to right: 
$t=1000,\; 4\times 10^4,\; 1.2 \times 10^5, 3\times 10^5$.
After an initial spreading time of approximately $4\times 10^4$ LB units, it takes roughly a time of $\bar{t}=8\times 10^4 \approx 3.6 \tcap$  for the passage of
the droplet over the step gradient.
}
\label{fig:lbm_spreading_on_step_grad}
\end{figure}}

Recalling the discussion of forces driving the motion of a spherical
cap on an open substrate (panel d) in \fref{fig:pL}), we
expect a slower separation dynamics for the \emph{same difference}
in contact angle in the hydrophobic regime. In order to examine this
expectation, we performed lattice Boltzmann simulations for the same
difference of contact angles $\theta_2-\theta_1=30^\circ$, but in
the more hydrophobic regime by choosing $\theta_1=105^\circ$ and
$\theta_2=135^\circ$ (while keeping all other parameters of the
simulation such as the droplet size, the system size, etc.\
unchanged). In agreement with our qualitative estimate, the time
necessary for the transfer of the whole droplet to the hydrophilic
part is significantly larger than for the case discussed in the
context of \fref{fig:lbm_spreading_on_step_grad}.

An important feature of computer simulation of fluid dynamical problems
is that not only one has access to the system density at each point in space but
also the whole velocity field is known. This is an interesting property
since it allows a survey of the liquid motion \emph{inside}
the droplet, whereby providing a means to estimate the viscous
dissipation. The latter, in turn, plays a crucial role in determining
the droplet dynamics.

Figure \ref{fig:vfield}
shows an example of this strength of computer simulations.
The left part of the figure illustrates a projection of the
velocity field onto the $xz$-plane for the case studied in
\fref{fig:lbm_spreading_on_step_grad} at a time of $t=10^4$
(LB unit) corresponding to $\that \equiv t / \tcap \approx 0.45$
(recall that $\tcap \approx 2.2\times 10^4$ for the present choice of
parameters).

In order to distinguish internal fluid velocity from the
velocity of the vapor phase, we also depict the liquid-vapor
interface in \fref{fig:vfield}. This allows
to recognize that the fastest lateral motion occurs
close to, but not exactly at, the three phase contact
line. During droplet spreading, the liquid-vapor interface
close to the hydrophilic substrate moves towards right. Quite
similarly but with a \emph{smaller magnitude}, the liquid-vapor
interface close to the hydrophobic substrate
moves towards left.

As a result of this asymmetric motion, the droplet spreads over the substrate while at the same time its
center of mass is shifted towards the more hydrophilic part of the substrate.

Obviously, the observed horizontal motion of the droplet's center of mass
is a result of the presence of a step gradient as
highlighted via a comparison to droplet spreading on a chemically
\emph{homogeneous} substrate (i.e.\ a substrate with a spatially
constant contact angle; see the right panel of
\fref{fig:vfield}). In this case, the droplet symmetrically
spreads on the substrate in order to reach its
static contact angle, the latter being an input parameter of
the simulation.

\hide{\begin{figure}
\epsfig{file=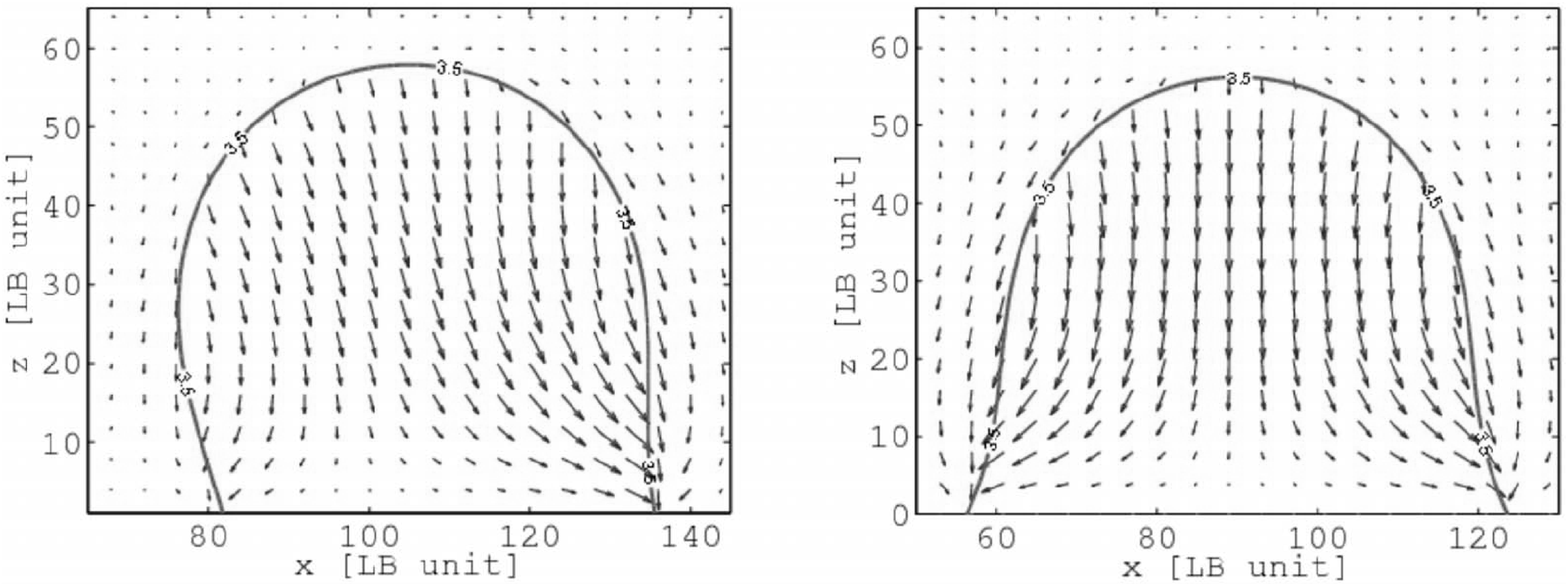,height=60mm,clip=}
\caption[]{Left: The velocity field within a $xz$-plane
at $y=L_y/2$ during the spreading of a droplet on a substrate with
a step-wise change in contact angle (changing from $\theta=105^\circ$
for $x<105$ to $\theta=75^\circ$ for $x \ge 105$). The solid line is the
liquid-vapor interface (the location of all
points with a density equal to
$(\rho_{\mathrm{gas}}+\rho_{\mathrm{liquid}})/2$).
Obviously, the presence of wetting gradient gives rise to an asymmetric
velocity distribution within the droplet.
Right: A similar plot
but for a droplet spreading on a chemically homogeneous substrate
(static contact angle $\theta=60^\circ$ for all $x$). As expected, both the shape
of the liquid-vapor interface and the velocity field inside the droplet are symmetric. All the snapshots correspond to a time of $t=10^4 \approx 0.45\tcap$ after deposing a spherical droplet of
radius $r_0=30$ lattice units on the substrate.
}
\label{fig:vfield}
\end{figure}
}

\subsubsection{Competitive effects of wetting gradient and flow pressure: The effect of droplet size}

The above lattice Boltzmann simulations address the behavior of a
droplet on a step gradient in the absence of flow, thus allowing to
focus on the effect of a step gradient alone. When flow is present,
there will be a possibility for an accidental deflection of the
polar component of the mixture (water in our experiments) towards
the more hydrophobic ('wrong') side. This may be caused e.g.\ due to
imperfect geometry of the channel giving rise to random fluctuations
in the velocity field.

As the deflected droplet moves over the step gradient (towards the
more hydrophobic side), a gradient of Laplace pressure is formed
inside the droplet pushing it back to the more hydrophilic part of
the substrate. In order for the separation process to be successful,
this wetting gradient induced force must overcome the forces driving
the flow. \blue{In a two component system, this force is usually  
the 'flow pressure'. Since our simulation studies are restricted to a 
one-component fluid and its vapor, we 'mimic' the flow pressure
via an external gravity-like force.}

Therefore, it is useful to consider what happens if a droplet is not only subject
to the effects of separation forces stemming from the wetting
gradient, but also to an opposite force. Figure
\ref{fig:sphericalcap_on_gradsurf} is devoted to such a situation.
To simplify the matter, the flow pressure is mimicked by an
external, gravity-like force \blue{density}. Each row
in \fref{fig:sphericalcap_on_gradsurf} illustrates a
spherical cap pushed under the action of exactly the same force
density towards a chemical step (from hydrophilic part towards
hydrophobic one).  As the droplet reaches the chemical step, the
corresponding gradient in Laplace pressure tries to hinder its
motion and to keep it on the more hydrophilic side.

Despite the fact that the both droplets are subject to exactly the
same force \blue{density}, the presence of a wetting gradient leads to
quite different dynamic behavior depending on the droplet size.
While the larger droplet passes under the action of the external
force over the step gradient (upper images in \fref{fig:sphericalcap_on_gradsurf}),
the smaller one is stopped at the
boundary between the hydrophilic and hydrophobic parts of the
substrate (lower images).

Following \eref{eq:fxcap}, this observation can be rationalized as follows.
The wettability gradient induced force scales as $1/\Omega^{2/3}$ and
thus as $1/r_0^2$, where $r_0$ is the radius of
the initial droplet on the substrate.
Indeed, the two droplets in \fref{fig:sphericalcap_on_gradsurf} differ by a
factor of 2 in their initial radii. The force opposing the droplet motion is,
therefore, expected to be by a factor of 4 larger in the case of
the smaller droplet as compared to the larger one.

As a first test of the scaling behavior given in \eref{eq:fxcap}, we increased the external
gravity-like force by a factor of four in order to see whether the smaller
droplet is pushed over the step gradient zone in a way similar to the motion
of the larger droplet. This naive expectation is,
however, not confirmed via our simulations. Nevertheless, an increase of the
external gravity like force by a factor of 5 turns out to be sufficient
for overcoming the step gradient effects.

Noting that the derivation of \eref{eq:fxcap} is based on a fully \emph{static} argument
(setting the fluid velocity $\bu=\vec{0}$ in the Navier-Stokes equation), the observed deviation between the
dynamic behavior and this estimate is not surprising. Rather, it emphasizes the fact that
a more quantitative analysis must involve the dynamics of the fluid within the droplet
and the corresponding dissipation losses. We are currently working on such a more elaborate
analysis.

It is instructive to also discuss various stages of droplet motion
for the case of a droplet being pushed over the step gradient
zone. Figure \ref{fig:dropletvelocity} shows the center of mass velocity
of a fluid droplet on a substrate consisting of a hydrophilic and a
hydrophobic part under the action of a gravity-like force. At time
$t=0$, a hemisphere of initial radius $r_0=40$ (LB units) is placed on the
hydrophilic part of the substrate and a gravity-like force \blue{density}
of $g=10^{-6}$(LB units) is switched on.
\hide{\begin{figure}
\epsfig{file=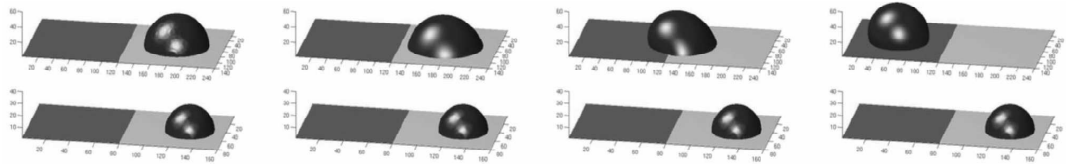,height=25mm,clip=}
\caption[]{Top: At time $t=0$, a hemisphere of radius $r_0=40$
(LB units) is placed on the hydrophilic (light (grey online); static contact angle
$\theta=75^\circ$) part of a substrate and a gravity-like force
of $g=10^{-7}$ (LB units) is switched on pushing the droplet towards the hydrophobic
part (dark (red online);  $\theta=105^\circ$). The time increases
from left to right as $t=0, \; 3\times 10^3,\; 10^6,\; 2.7\times
10^6$ (LB units). Bottom:  The same situation as in the upper
panels, but for a smaller droplet of $r_0=20$ LB units. In  this case, gradient
of Laplace pressure over the chemical step is strong enough in order
to fully stop the droplet motion (from left to right: $t=0,\;
2\times 10^5,\; 5 \times 10^5,\; 1.5\times 10^6$).}
\label{fig:sphericalcap_on_gradsurf}
\end{figure}}

The lateral extension of the hydrophilic side of the substrate is
chosen such that the droplet can reach its steady state motion
before arriving at the interface between the hydrophilic and
hydrophobic parts of the substrate (first plateau in the velocity in
\fref{fig:dropletvelocity}). At the wettability step, the
droplet velocity decreases indicative of additional (wetting
gradient induced) forces opposing its motion. The droplet velocity
reaches a minimum and then increases again towards a new plateau
corresponding to steady state motion on the hydrophobic part of the
substrate.

Note that, in agreement with various experimental observations as
well as computer simulations \cite{Barrat1999}, we observe a higher
steady state droplet velocity over a hydrophobic substrate as
compared to the hydrophilic one under the action of exactly the same
external force. This is indicative of lower dissipation losses for a
motion on hydrophobic substrates.

As a consequence of the periodic boundary condition present in our
simulations, we can also survey what happens when the external
driving and wetting gradient forces act along the \emph{same}
direction. This occurs as the droplet leaves the channel at the left
boundary and reenters it from the right side, thereby leaving
the hydrophobic substrate towards the hydrophilic one. In this case,
the droplet motion is accelerated by the action of the wetting
gradient (see the maximum in \fref{fig:dropletvelocity}).
Finally, as the gradient zone is left behind, the droplet
decelerates to reach its steady state velocity on the hydrophilic
substrate.

\hide{\begin{figure}
\epsfig{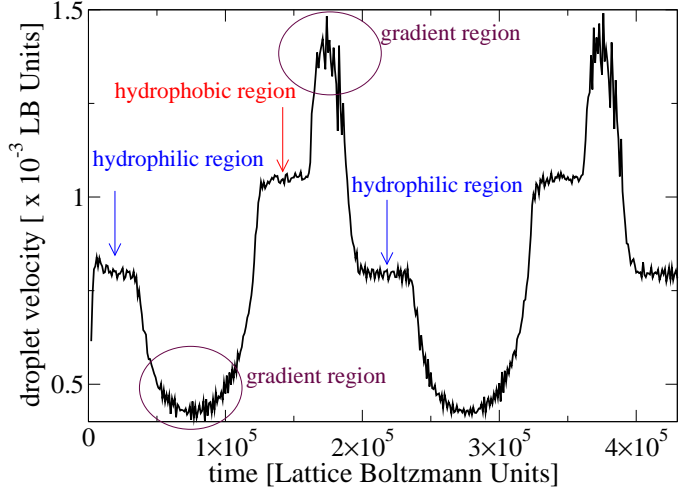}
\caption[]{(Color online) Different stages during the motion of a fluid
  droplet (hemisphere) \blue{of radius $r_0=20$ lattice units under the action of an
  external force \blue{density} of $g=10^{-6}$ (LB units)} on an open substrate made
  of a periodic array of hydrophilic \blue{(light (grey online); $\theta=75^\circ$)} and
  hydrophobic \blue{(dark (red online);  $\theta=105^\circ$)} parts. \blue{The system size is
  $160 \times 80\times 40$ (along $x$, $y$ and $z$-directions). Other
  parameters of the simulation are $\tau=0.8$, $\kappa=0.002$ and $T=0.4$.}}
\label{fig:dropletvelocity} 
\end{figure}}

Different stages of the droplet motion over a step gradient are also
illustrated in \fref{fig:sphericalcap_on_gradsurfB} for a smaller droplet
(radius $r_0=20$ lattice units) moving under the action of a driving force
of $g=10^{-6}$ [LB units]. In this figure, we focus
on the droplet shape and its variation as the droplet passes over
the wettability step. As expected, the droplet spreads better on the hydrophilic part of
the substrate than on the hydrophobic part. As a consequence, its lateral
extension decreases leading to an increase in its height during the passage
over the step gradient zone (moving from right to left).

Note that the shape of a droplet moving on a substrate depends on its velocity.
A constant droplet shape may, therefore, give a hint on the presence of a
steady state motion. Keeping this in mind, a closer look at regions sufficiently
far from the wetting gradient zone in
\fref{fig:sphericalcap_on_gradsurfB} points to such a steady state motion on both
hydrophilic and hydrophobic parts of the substrate. A survey of the velocity
versus time (similar to that shown in \fref{fig:dropletvelocity}) confirms this
observation.

\hide{\begin{figure}
\epsfig{file=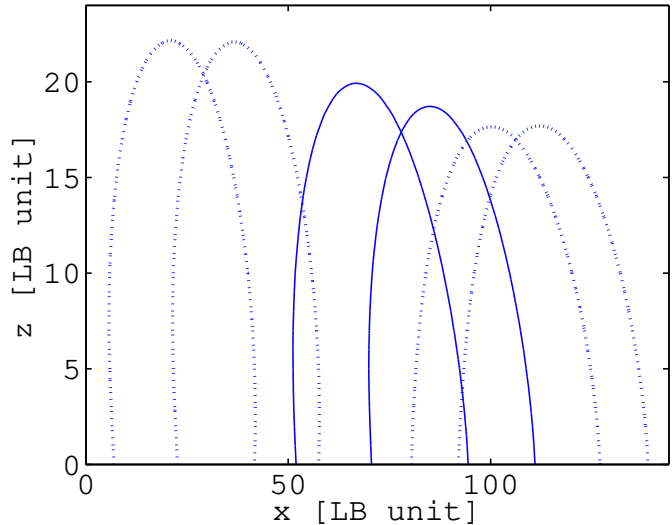,height=75mm,clip=}
\caption[]{(Color online)  A cut along the $xz$-plane of the liquid-vapor
  phase boundary at $y=L_y/2=40$ for a spherical cap of initial radius
  $r_0=20$ (LB units). The interface line is shown at different times during
  the motion of the spherical cap upon the action of a gravity-like force \blue{density} of $g=10^{-6}$ (10 times larger than the force applied in the case of \fref{fig:sphericalcap_on_gradsurf}). The time increases from right to left as $t = 2,\; 3.5,\; 6,\; 10,\; 14,\; 15.5$ [$\times 10^4$ LB units]. The dotted lines correspond to the steady state motion on the hydrophilic (right) and hydrophobic (left) parts of the substrate, whereas the solid lines  correspond to the transition over the wettability step (placed at $x=L_x/2=80$).}
\label{fig:sphericalcap_on_gradsurfB}
\end{figure}}

We also studied the competing effects of a wetting gradient and an
external driving force  in
the case of a slit geometry
(\fref{fig:liquidfilm_on_gradsurf}). Here, we focus on the effect of
channel height on the force balance.
For this purpose, all parameters of the simulation are kept exactly the same.
Only the vertical separation between the substrates is varied.
Note that, since we keep the lateral extension of the fluid inside the slit constant,
the wettability induced driving force varies as
$\fxdriv \propto \delta \pL/l_x \propto \delta \pL \propto 1/H$,
where \eref{eq:pLslit} for $\pL$ in a planar slit is used.
As shown in \fref{fig:liquidfilm_on_gradsurf}, our simulation
results are, at least qualitatively, in line with the expected
reduction of $\fxdriv$ upon an increase of the channel height.

\hide{\begin{figure}
\epsfig{file=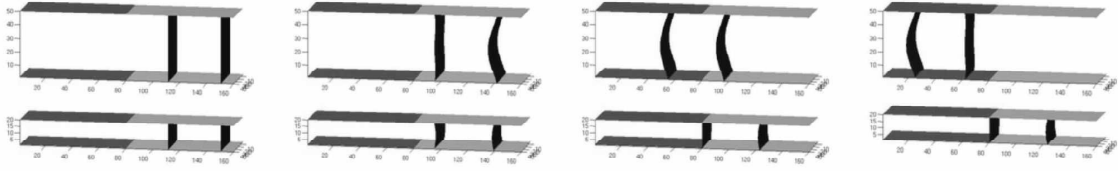,height=25mm,clip=}
\caption[]{Top: Snapshots of a fluid film in a planar slit of height
$H=50$ (LB units) as it is pushed by the action of an external
gravity-like force towards left, passing thereby from the
hydrophilic (gray; static contact angle $\theta=75^\circ$) part of
the substrate to  the hydrophobic part (dark (red online);  $\theta=105^\circ$).
At time $t=0$ (the image on the left),
the space delimited by the two planar interfaces is
filled with liquid (for clarity, not the
liquid but only the liquid/vapor interfaces are shown). The other
three panels correspond  to (from left to right) the steady
state motion over the hydrophilic substrate ($t=10^4$), motion over
the step gradient ($t=4\times 10^4$) and the steady state motion
over the hydrophobic part of the substrate ($t=6\times 10^4$). Bottom:
The same situation as in the upper  panels, but for a smaller
channel height of $H=20$ (LB units). In this case, wetting gradient
induced forces are strong enough in order to fully stop the fluid
motion (from left to right $t=0,\; 5\times 10^4,\; 2\times 10^5,\;
5\times 10^5$). In the both studied cases, the external force
density  is  $g=10^{-6}$ (LB units).}
\label{fig:liquidfilm_on_gradsurf}
\end{figure}}

\section{Summary and Conclusion}
Basic aspects related to the wetting gradient induced separation of
a binary emulsion into its individual components are studied
via experiments as well as lattice Boltzmann computer simulations.
For the purpose of experimental investigations, an emulsion
generator is designed allowing the production of quasi-monodisperse
emulsions with well controlled droplet sizes
(\fref{fig:emulsion-generation}).

As to the wetting gradient, a
step-like change in wetting properties of the substrate is realized
via dip-coating process (\fref{fig:dipcoating}). Compared to a
substrate with a continuous change in wetting properties, the
step-like gradients have the advantage of e.g.\ easy fabrication and
long time storage possibility allowing their use up to many weeks
after fabrication. Basic results of our separation experiments can
be summarized as follows.

It is observed that separation in confined geometry is far more
enhanced than on open substrates. Furthermore, the separation
process is mainly driven by the selectivity of the water component,
i.e.\ by the tendency of water to preferentially wet the more
hydrophilic side of the substrate. This is best seen in a closed
channel: Once water fills the whole cross section of the separation
chamber, it pushes the other component of the mixture (toluene or
hexane in our experiments) to the less hydrophilic part of the
channel (\fref{fig:sep_exp2}).

This property is then used in
order to enhance the separation efficiency. For this purpose, a
second inlet is introduced on the opposite side of the channel
through which water is injected into the separation chamber
(\fref{fig:sep_exp3}). The separation starts as soon as the two water
streams meet in the gap and build a water front spanning the whole
cross section of the channel.

These experimental observations are accompanied with order of
magnitude estimates of the dominant forces responsible for
separation [Eqs.\ (\ref{eq:pLcap})-(\ref{eq:fxslit}) and \fref{fig:pL}].
Even though neglecting many complexities present in the real problem, these estimates
provide at least a qualitative understanding of the significantly
different behavior of wettability induced forces upon a variation
of the contact angle in a narrow slit as compared to open substrates.

The analytic estimates given in section \ref{subsection:drivingforce}
are in qualitative agreement with lattice Boltzmann
computer simulations (section \ref{subsection:simulation}).
In particular, on open substrate,
a stronger gradient in Laplace pressure is observed for the same
difference in contact angle in the hydrophilic regime
rather than in the hydrophobic one. Such an effect is fully absent in the case of
a planar slit, since the dependence of $\pL$
on the contact angle is symmetric upon a variation
of $\theta$ around $\theta=90^\circ$ [\fref{fig:pL}].

Using lattice Boltzmann simulations, we also investigate
the effect of a step-like wetting gradient on the motion of
a droplet placed at the gradient zone in the absence of flow
(Figs.\ \ref{fig:lbm_spreading_on_step_grad} and \ref{fig:vfield}).
These simulations confirm that a single step-wise change in the
contact angle is sufficient in order to induce a preferential deflection of
the droplet towards the region of lower contact angle
(\fref{fig:lbm_spreading_on_step_grad}).

Via simple scaling arguments we also give an approximate expression for
the characteristic time, $\bar{t}$, for the passage of a droplet over the wettability
step in the absence of external forces. This time scale turns out to
depend on the capillary time, the initial droplet volume as well as
on the static contact angles of the hydrophilic and hydrophobic parts
of the substrate [\eref{eq:tav}]. Lattice Boltzmann simulations
show that the expression given in \eref{eq:tav} yields the correct
order of magnitude estimate of $\bar{t}$.

Moreover, a study of the internal fluid dynamics yields interesting
insight on how the fluid redistributes within the droplet in order
to accommodate both the spreading dynamics and the lateral motion of
the droplet towards the more hydrophilic part (\fref{fig:vfield}).

It would be interesting to examine these computer simulation results
in the light of experimental observations. Owning to considerable
difficulties in the observation of the local fluid motion, such
experiments have not been available yet. Nevertheless, we are
planing the use of tracers for visualization purpose and high speed
cameras allowing to resolve the fast spreading dynamics of a fluid
droplet on a step gradient.

Furthermore, the important role of droplet size as well as channel dimension
for the separation process is underlined via analytic estimates
[Eqs.\ (\ref{eq:pLcap})-(\ref{eq:fxslit})]
and computer simulations
[Figs.\ \ref{fig:sphericalcap_on_gradsurf} and \ref{fig:liquidfilm_on_gradsurf}].
While the motion of smaller
droplets is dominated by the action of wetting gradient induced
forces, larger droplets tend to follow the external gravity-like
force [\fref{fig:sphericalcap_on_gradsurf}].
A similar effect occurs in the
case of a narrow slit (\fref{fig:liquidfilm_on_gradsurf}).
Here, the effect of wetting gradient induced forces is weakened by
an increase of the channel height (vertical separation between the
substrates) thus further emphasizing the importance of
surface/volume ratio for the separation efficiency.

An important conclusion of our studies is that, both on
open substrates and in closed planar channels, larger droplets
are harder to guide by a step-wise wetting gradient than are
the smaller ones. In the both studied geometries, the wetting
gradient induced force decreases with droplet volume
$\Omega$ (scaling as $1/\Omega^{2/3}$ on open substrates
and as $1/(H\Omega)^{1/2}$ in a planar slit; $H$= channel height).

Since the smallest droplet size is controlled by the dimensions
of the emulsion generator (see e.g.\ \fref{fig:emulsion-generation})
as well as by the height of the planar slit, these observations
suggest that a miniaturization of the channel would allow the
generation and use of smaller droplets and thus could lead to a
significant improvement of the separation efficiency.

Even though of a rather qualitative nature, results presented in
this report shed light to some of the basic issues related to the
separation phenomena, whereby allowing to isolate interesting
aspects for a more quantitative analysis. Such an investigation is
the subject of ongoing work.

\vspace*{2mm}
\noindent{\bf\large Acknowledgments}

This project was supported by the German Research Foundation
(DFG) within the priority program SPP1164, Nano-and Microfluidics,
project numbers  Va205/3-2 and Sta324/27-2. FV thanks N.\ Peranio
for his assistance during the preparation of this manuscript
and A.\ Dupuis for providing the 3D LB code, which we thoroughly
tested and slightly modified for our own purposes.


\end{document}